\documentclass{article}

\usepackage[dvips]{hyperref} 
\usepackage[dvips]{graphicx}
\usepackage{amssymb,amsfonts,amsmath}

\begin{document}


\centerline{\bf\Large Neuropsychological constraints to}
\centerline{\bf\Large human data production on a global scale}

\bigskip \bigskip

\centerline{Claudius Gros$^{1}$, Gregor Kaczor$^{1}$,
            Dimitrije Markovi\'c$^{1}$}
\bigskip

\noindent
$^1$Institute for Theoretical Physics, Goethe University
   Frankfurt, Germany.
\bigskip \bigskip


\begin{abstract}
\end{abstract}
\section*{Abstract}
Which are the factors underlying human information
production on a global level? In order to gain an
insight into this question we study a corpus of 252-633 mil.\
publicly available data files on the Internet
corresponding to an overall storage volume of
284-675 Terabytes. Analyzing the file size distribution
for several distinct data types we find indications that
the neuropsychological capacity of the human brain
to process and record information may constitute the dominant
limiting factor for the overall growth of globally stored
information, with real-world economic constraints having
only a negligible influence. This supposition draws support
from the observation that the files size distributions follow
a power law for data without a time component, like images,
and a log-normal distribution for multimedia files, for
which time is a defining qualia.

\section*{Author Summary}
The generation of new information is limited by two key
factors, by the incurring economic costs and by the capacity 
of the human brain to process and store data and information; 
the controlling agent needs to retain an overall understanding 
even when data is generated by semi-automatic processes. 
These processes are reflected in the statistical properties 
of the data files publicly available on the Internet. 
Collecting a corpus of 252-633 mil.\ files we find that
the statistics of the file size distribution are consistent
with the supposition that data production on a global level 
is shaped and limited by the neuropsychological information 
processing capacity of the brain, with economic and hardware 
constraints having a negligible influence.

\section*{Introduction}
Information production and storage becomes
progressively easier. Moore's law \cite{moore98} states 
that technological advancements lead to a doubling of
computing power every 1.5 years and that
data storage capacity increases by a factor of 
about 100 every 10 years \cite{gray02}.
Data production, which has the goal to 
increase knowledge and information, is
constrained on one side by the economic costs involved
and on the other side by the neuropsychological 
limitations and costs of the data generating agents. 
Maximizing the total amount of information generated 
for given amounts of economic and neuropsychological 
resources hence determines the shape of the file-size 
distribution \cite{gros10}.

The economic costs for data production involving
hardware, software and management are
proportional to the amount of data produced.
The overall goal of data production is the
generation of information, which can be measured
by Shannon's information entropy \cite{shannon49}.
Maximization of information entropy under the
constraint of economic costs leads to
file size distributions having exponential
tails \cite{gros10, markovic10, mandel53}. 
Exponential tails are however absent both 
in our data and in an earlier 
study of the file-size distribution
on a large number of Windows file systems on desktop 
computers \cite{douceur99}. The absence of exponential 
tails for files hosted on Internet servers indicates 
that economic costs are not the limiting factors for 
data production.

The ability of the human brain to process and record 
information determines a subjective value which 
the producing individual attributes to an 
information source. E.g.\ the amount
of information gained when increasing the resolution of 
a low quality image is substantially higher then when
increasing the resolution of a high quality photo by 
the same degree. This relation is known as Weber-Fechner 
law and results from underlying neurophysiological 
processes \cite{hecht24,nieder03,dehaene03}.
We find that the observed file-size 
distributions on the Internet are consistent with 
the Weber-Fechner law and propose that 
neuropsychological constraints may be a dominant
factor in shaping the statistics of global 
data production.
This hypothesis is based on the finding
that the distribution functions maximizing
information entropy, given the neurophysiological
constraints of the Weber-Fechner law, nicely 
reproduce the real world file-size distributions.

The neurophysiological constraints resulting
from the  Weber-Fechner law also imply that
the different maximal entropy distributions 
are qualitatively different for data formats
involving time, like audio and video, compared
to file types not involving time, as it is the case 
for images. We find that these distinct predictions 
are very well in agreement with the observed 
files-size distributions.

\begin{figure}[t]
\centerline{
\includegraphics[width=0.8\textwidth]{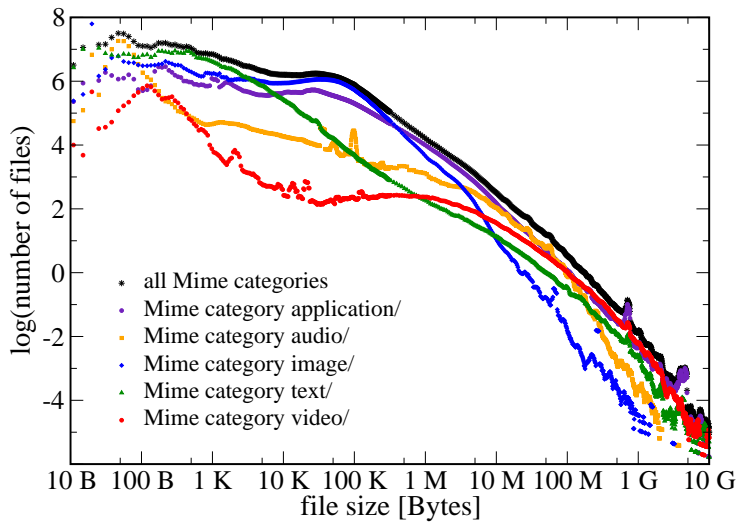}
           }
\caption{{\bf File-size distribution for 252 mil.\ files
         hosted in 7.7 domains.}
For all files types together and for the five Mime categories
individually. Displayed is the log$_{10}$ of the number of files in
bins of 1 Kbyte.}
\label{fig_all}
\end{figure}

\subsubsection*{Maximal Entropy Distribution Functions}
Given a normalized distribution function
$P(s)$ for a corpus of data, in our case the 
file-size distribution, its information content 
can be measured by Shannon's information 
entropy \cite{shannon49}, $-\sum_s P(s)\log(P(s))$.
The overall goal of data production, to attain an 
optimal information content, is achieved when 
the respective information entropy is maximal. 

We denote with $c(s)$ the cost function associated
with the economic and neurophysiological constrains.
and with $\lambda$ the respective Lagrange
multiplier. The distribution function $P(s)$ 
maximizing information entropy given the
constraint $c(s)$ is determined by \cite{gros10, mandel53}
\begin{equation}\label{eq_delta}
\delta \Lambda[P] \ =\ 0, \qquad\quad
\Lambda[P] = \sum_s P(s)\log(P(s))-\lambda\sum_s P(s)c(s)~,
\end{equation}
where $\delta \Lambda[P]$ denotes the variation of the
functional $\Lambda[P]$ with respect to distribution
functions $P(s)$. One obtains from (\ref{eq_delta})
that $P(s)\sim\exp(-\lambda c(s))$. The maximal entropy 
distributions have then the form
\begin{equation}\label{eq_distr}
P(s)\sim\exp\big[-\lambda_s s-\lambda_1\log(s)-\lambda_2\log^2(s)\big]~,
\end{equation}
when considering cost functions containing terms
proportional to the files size $s$,
to $\log(s)$ and to $\log^2(s)$. The first term,
linear in the size of the files $s$, corresponds to
economic costs. The other two terms in the cost functions 
correspond to the scaling of neurophysiological resources.

The Weber-Fechner relations state that
the neural representations of sensory 
stimuli \cite{hecht24}, objects 
\cite{nieder03,dehaene03,nieder02}
and time perception \cite{takahashi05} in the brain
scale logarithmically with the intensity of the stimuli,
the number of objects and the length of the
time interval respectively. The perceived costs 
and benefits of information generation and processing 
hence scale logarithmically with physical data volume.
Maximization of information entropy under the logarithmic
cost function yields a power-law file size distribution,
as described by Eq.\ \ref{eq_distr}.

The perceived cost function will scale furthermore 
as the square of the logarithm whenever the data is 
characterized by two neurophysiological distinct degrees 
of freedom, such as resolution and time.
The distribution maximizing information entropy will
then be a log-normal file size distribution,
see Eq.\ \ref{eq_distr}. We find that this is indeed 
the case for multimedia files, such as audio and video 
files, for which the time is defining qualia. 
The file size distributions of non-temporal data 
types (e.g.\ texts and images) follow, on the other
side, a power-law. 

If the cost function scales as the square of the logarithm, 
the file-size distribution maximizing information entropy 
will then have a log-normal form, see Eq.\ \ref{eq_distr}.
We find that this is indeed the case for multimedia files, 
such as audio and video, for which the time is defining qualia.
The file size distributions of non-temporal data types 
(e.g.\ texts and images) is closer, on the other side, to 
a power-law.

\begin{figure}
\centerline{
\includegraphics[width=0.8\textwidth]{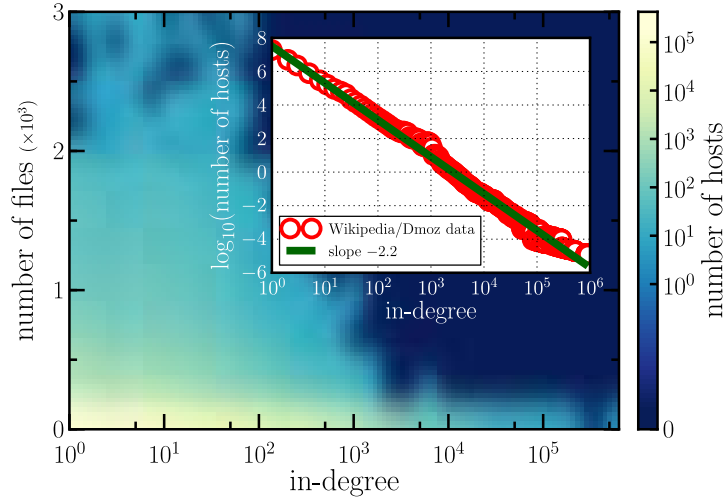}
           }
\caption{{\bf File hosting vs.\ in-degree.}
Main: The number of domains (dark blue: few hosts,
white: many hosts) with a given in-degree (x-axis)
and hosting a given number of files (y-axis);
all Mime categories without text/ and image/.
Inset: For the 32 mil.\ hosts receiving incoming links from
the Wikipedia/Dmoz corpus the distribution of the
in-degree.}
\label{fig_degree}
\end{figure}

\section*{Results}
We performed a large scale search of publicly
available files on the Internet, utilizing the
spider of file search engine FindFiles.net. For
the corpus of hosts to be crawled we selected
the collection of all outgoing links in Wikipedia.org
and Dmoz.org, the open directory project, scanning
in both cases all available editions. We crawled,
in a first effort, a total of 7.7 mil.\ hosts, indexing
252 mil.\ data files. The resulting file size
distribution is presented in Fig.\ \ref{fig_all} 
in a log-log representation, spanning nine orders 
of magnitude. The crawling effort was then continued
in a second stage until a corpus of 633 mil.\ files
had been reached, which we used for a systematic study
of the statistical properties of the resulting
file-size distribution.

\subsubsection*{File Taxonomy}
Data files can be classified 
according to their Mime or Internet Media 
Types, e.g.\ a jpeg file has the Mime type
{\sl image/jpeg} within the Mime category {\sl image/}.
Five Mime categories make up about 99.9\% of all data formats
publicly accessible on the Internet, with {\sl application/}
contributing 33.2\%, {\sl audio/} 2.9\%, {\sl image/} 58.0\%,
{\sl text/} 5.1\% and {\sl video/} 0.7\% respectively to the total number
of files in the Wikipedia/Dmoz corpus. The average number of
files per host, the average file sizes (in Kbytes) and
median file sizes (in Kbytes) are respectively
(10.8\textbar1312\textbar136) for {\sl application/},
(0.9\textbar6733\textbar1589) for {\sl audio/},
(19.0\textbar189\textbar72) for {\sl image/},
(1.7\textbar3786\textbar5) for {\sl text/} and
(0.2\textbar28912\textbar5548) for {\sl video/}. The average
file size of 189 Kbytes for images in our data has seen an
increase relative to the 15 Kbytes found in an earlier
study \cite{lawrence99}. The substantial difference
between the respective means and medians is a consequence
of the fat tails in the corresponding distribution functions,
compare Fig.~\ref{fig_all}.

\begin{figure}
\centerline{
\includegraphics[width=0.8\textwidth]{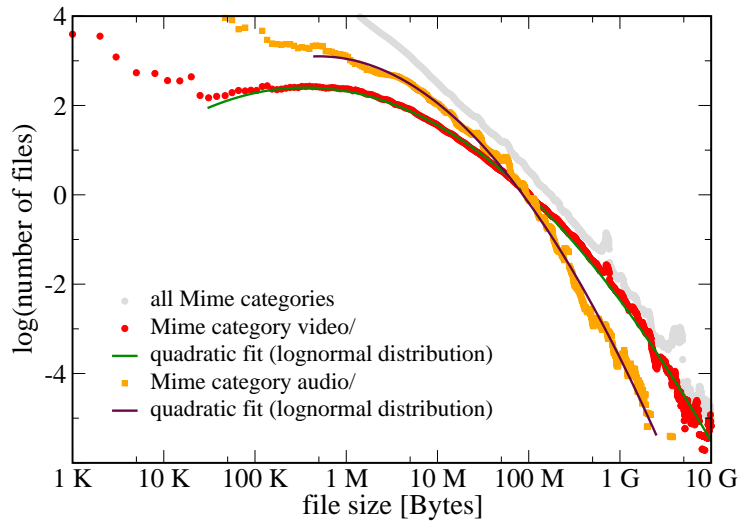}
           }
\caption{{\bf File-size distribution for videos and audio
files (252 mil.\ files).} 
Solid lines represent an eye guide of a quadratic form, 
$a*\log(\mathrm{size}) - b*\log^2(\mathrm{size})$, 
where $a,b > 0$. }
\label{fig_video}
\end{figure}

\subsubsection*{File size distribution}
Fig.\ \ref{fig_degree}
shows the correlation between the number of files hosted
and the in-degree (the number of inbound links) of the
hosting domain. Important domains tend to have a high 
in-degree \cite{boccaletti06}, e.g.\ the in-degree of 
Twitter.com is 805000 in the Wikipedia\-/\-Dmoz corpus. The number of
publicly accessible data files hosted is however
anti-correlated with the in-degree, most data
being hosted on relatively unknown hosts. The power-law
for the in-degree distribution presented in the inset of
Fig.\ \ref{fig_degree} has remained remarkably constant for
the World Wide Web over the last decades. Our slope of $-2.2$
for the 32 mil.\ hosts within an one-click distance of the
Wikipedia\-/\-Dmoz corpus is very close to the slopes
between $-1.94$ and $-2.1$ found in previous 
studies \cite{albert99,adamic00,kong08}.

A manifest property of the file size distribution
presented in Fig.~\ref{fig_all} is the absence
of exponential tails, which one would have
expected \cite{gros10,markovic10} for an information
entropy production constraint by economic limitations,
like costs and available storage space.
The lack of exponential tails has been observed in an earlier
study of the file size distribution on a large number of
Windows file systems on desktop computers \cite{douceur99}.
They have also found that the utilization ratio of
desktop hard disks is, on the average, below the capacity.
Thus, the full storage volume is rarely utilized by the
average PC user. 

\begin{figure}
\centerline{
\includegraphics[width=0.8\textwidth]{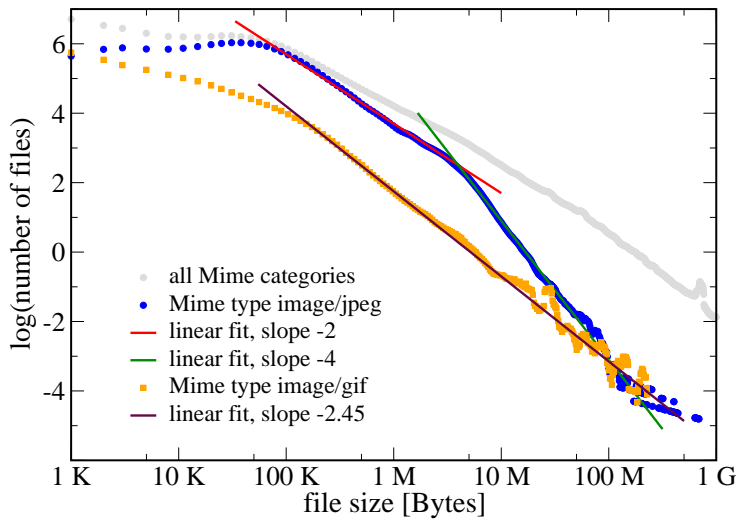}
           }
\caption{{\bf File-size distribution for jpeg and gif images
(252 mil.\ files).}
The transition from a -2 to a -4 slope for the jpeg-file
distribution occurs at about 4 Mbyte. This kink can be
attributed to the transition from amateur to professional
image production.}
\label{fig_image}
\end{figure}

A differentiated perspective can be obtained when
examining the functional form of the file size distributions
for distinct Mime categories and types. The
tails for the video and audio file distributions,
shown in Fig.\ \ref{fig_video}, and the tails
for the file size distributions of jpeg and gif
images presented in Fig.\ \ref{fig_image} differ
manifestly.

The linear dependence observed in Fig.\ \ref{fig_image}
corresponds to a scale-free power-law $P(s)\propto s^{-\gamma}$
of the file size distribution $P(s)$ with distinct slopes for
lossless and lossy image compression formats, gif and jpeg,
respectively. For video and audio files the file size
distribution follows a log-normal dependence, with
$\log(P(s))\propto \alpha\log(s)-\beta\log^2(s)$
fitting the data excellently over more than
5 orders of magnitude. These two distributions
differ qualitatively in two aspects, namely in
the occurrence of the quadratic term $\log^2(s)$ for the
log-normal distribution and in the sign of the
linear term. The leading term $-\gamma\log(s)$
has a negative slope for image data formats and a
positive slope $\alpha\log(s)$ for the {\sl audio/} and
{\sl video/} Mime categories (with $\alpha,\gamma >0$).
The log-normal dependence observed for
video and audio files is hence qualitatively
distinct with respect to a power-law scaling
and cannot be interpreted as a quadratic
correction to a linear fit within a log-log data analysis.

The fact that the file size distributions and the distribution 
tails are qualitatively different for multimedia
and image file formats, strongly indicates that 
they are determined by the underlying neurophysiological 
limitations of the producing agents. The cost functions 
are therefore, see Eq.\ (\ref{eq_distr}), proportional to
$\log(s)$ and $\log^2(s)$ for data characterized by 
one and two degrees of freedom, respectively.

\begin{figure}
\centerline{
\includegraphics[width=0.8\textwidth]{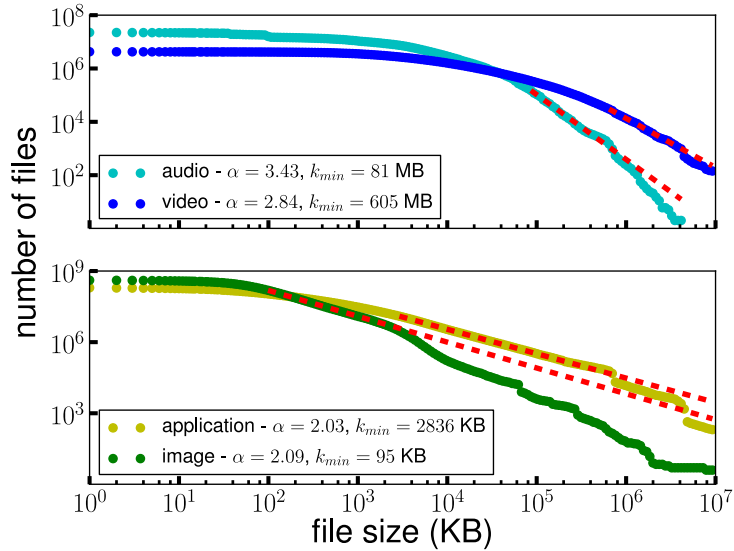}
           }
\caption{{\bf Power-law fit to the complementary 
cumulative file-size distribution.} The red dashed 
lines are the fits, the respective parameters 
are given in the insets. For Mime categories
{\sl audio/}, {\sl video/} (top) having a time
component and {\sl application/}, 
{\sl image/} (bottom) having no time component.}
\label{powers}
\end{figure}
 
\section*{Fitting methods}

The guide-to-the-eye fits shown in 
Figs.\ \ref{fig_video} and \ref{fig_image} 
indicate that the statistical properties
of the file-size distributions depend on the
presence/absence of a time-component.

For a systematic analysis we used the corpus
of 633 mil.\ files, containing four Mime 
categories, {\sl image/} (64.8\%), {\sl application/} (31\%), 
{\sl audio/} (3.5\%) and {\sl video/} (0.7\%). 
The evaluation of file size distribution 
was performed in two steps. In a first step 
the files were binned into 1 Kbyte bins. 
In a second step we evaluated maximum likelihood 
estimates for two model distributions \cite{clauset09}.

\begin{figure}
\centerline{
\includegraphics[width=0.8\textwidth]{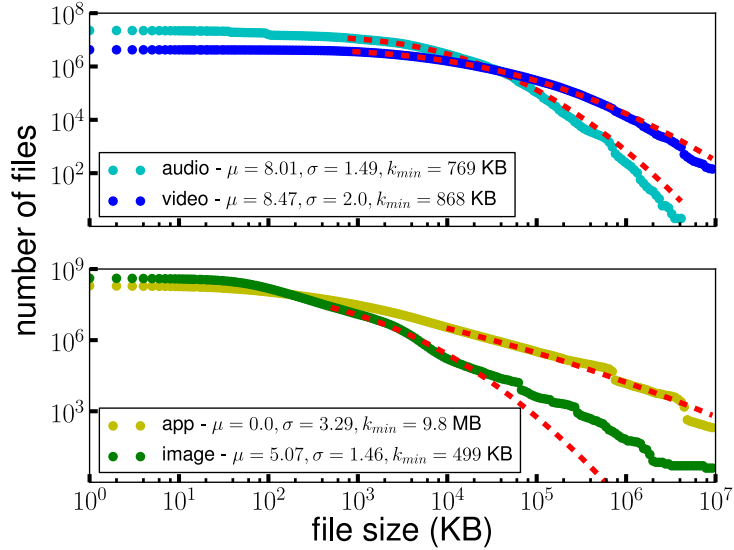}
           }
\caption{{\bf Log-normal fit to the complementary
cumulative file-size distribution.} The red dashed 
lines are the fits, the respective parameters 
are given in the insets. For Mime categories
{\sl audio/}, {\sl video/} (top) having a time
component and {\sl application/}, 
{\sl image/} (bottom) having no time component.}
\label{lognormals}
\end{figure}

We analyzed the tails of the respective
file size distributions with two
types of discrete probability distributions, 
a power law,
$$
p(k)\ =\ \frac{1}{Z_{k_{min},\alpha}}\, k^{-\alpha},
\qquad\quad
Z_{k_{min},\alpha} = \sum_{k_{min}}^\infty k^{-\alpha}~,
$$
and one having a log-normal form
$$
p(k)\ =\ \frac{1}{Z_{k_{min},\mu, \sigma}}
\,\frac{1}{k}\, e^{-\frac{(\log{k}-\mu)^2}{2\sigma^2}},
\qquad\quad
Z_{k_{min},\mu, \sigma} = 
 \sum_{k_{min}}^\infty \frac{1}{k}\,
e^{-\frac{(\log{k}-\mu)^2}{2\sigma^2}}~.
$$
In both cases a lower bound 
$k_{min}$ is introduced as a free parameter, 
as we don't expect describing the whole range 
of data but only the tails of available 
data by either a power-law or log-normal 
distribution.

The actual fitting procedure consist of following steps:
\begin{itemize}
\item We've first performed a maximum likelihood 
estimate for the lower bound $k_{min}$ in the 
range from 1 KB up to 100 MB.

\item Then, we have selected a $k_{min}$ which minimizes 
residual sum of squares ($rss$) of the differences
between the empirical and the fitted tails of the
complementary cumulative distribution functions, that is 
\begin{equation}
rss\ =\ \sum_{k'=k_{min}}^{k_{max}}
\big(\,\mathrm{Pr}(k \geq k^\prime)-F(k')\,\big)^2~,
\label{eq_rss}
\end{equation}
with $F(k) = \sum_{k^\prime= k}^\infty p(k')$ being the
complementary cumulative distribution of the model and
and $\mathrm{Pr}(k \geq k^\prime)$ respectively the
empirical complementary cumulative file distribution.
\end{itemize}
We present the best fits for the four Mime 
categories ({\sl image/}, {\sl application/}, 
{\sl audio/} and {\sl video/}) 
for a power-law distribution in Fig.~\ref{powers}, 
and a log-normal distribution in Fig.~\ref{lognormals}, 
respectively. In obtaining the maximum 
likelihood estimate for model parameters 
we have neglected files larger then 10 
Gbytes, as there are only very few of
these extremely big files, they are hence
statistically not representative.

Comparing the two fits for {\sl audio/}
and {\sl video/} data we find that the
log-normal distribution describes the empirical
data substantially better. The $rss$ values
are order of magnitude lower in the case of
log-normal fit (see Table~\ref{tbl_rss}).
The log-normal fit is also able to describe a broader
range of the data then the power-law fit 
(compare Fig.~\ref{powers} and Fig.~\ref{lognormals}).

Similarly, a power-law fit for {\sl application/}
file-size distribution describes a broader range
of the empirical data, and has an order of magnitude
smaller value, then the one obtained
for a log-normal fit (Table~\ref{tbl_rss}). In the
case of the {\sl image} file types the evidence in favor
of a power-law distribution is not particularly
strong, a consequence of the kink at around
4 Mbytes, compare Fig.~\ref{fig_image}.
Both fits, log-normal and power-law, describe
a similar data range and the corresponding $rss$ values
are of similar magnitude.

\begin{table}[b]
\caption{ Residual sum of squares, $rss$, estimated
as sum of square differences between empirical and fitted
complementary cumulative distributions (see Eq.~\ref{eq_rss}).}
\vspace{0.2cm}
\centerline{
\begin{tabular}{|c|c|c|}
\hline
  & \ power-law fit \ & \ log-normal fit \ \\
\hline
 \ {\sl image/} \ & \ $0.7$ \ & \ $1.0$ \ \\
\hline
 \ {\sl application/}  \ & \ $3.3$ \ & \ $47.9$ \ \\
\hline
 \ {\sl audio/} \ & \ $26.4$ \ & \ $2.1$ \ \\
\hline
 \ {\sl video/} \ & \ $451.4$ \ & \ $2.7$ \ \\
\hline
\end{tabular}
\label{tbl_rss}
}
\end{table}

\section*{Discussion}
For images the production costs are functionally
dependent on one variable, the resolution, which
defines, modulo compression algorithms, the file size.
The cost function for the production of videos depends however on
two distinct quantities, the resolution per frame and
the total number of frames, viz the time needed to
shoot the sequence. Analogously for audio files,
with frequency resolution and length being the
two cost defining quantities. The cost functions associated
with information production are hence one- and
two-dimensional for images and audio/video formats
respectively. We generically observe in our data
that one-dimensional cost functions
result in power-law file size distributions,
two-dimensional cost functions in log-normal
distributions. 

For compound Mime categories or types,
like {\sl text/}, superpositions of these two basic
distributions are observed. This correlation between
dimensionality of data type and resulting
file size distribution, which can
be seen in Fig.~\ref{powers} and Fig.~\ref{lognormals},
finds a straightforward rationale when accounting for
the neuropsychological constraints for data processing.

Our analysis is based on the assumption
that an ensemble average over many information producing
agents reveals the underlying information theoretical
principles driving data production on a global level.
Other studies have investigated alternative approaches, 
like the study of microscopic models capable of 
generating distributions with large tails, such
as scale-free \cite{barabasi99,gabaix03}
and log-normal \cite{mitzenmacher04,mitzen04b} 
and the double Pareto-lognormal distribution \cite{reed04}.
In a related context a log-normal distribution has
been found for the distribution of city sized and 
be related to proporionate growth mechanisms \cite{eckhout04}.

Both approaches, the modelling of generative processes
and the information theoretical perspective,
are complementary and do not exclude each other.
Ultimately it may be possible to derive classes of
microscopic generative models from comprehensive information
theoretical principles, as it has been proposed, e.g.,
for intrinsic neural adaption rules generating
information entropy maximizing firing rate
distributions \cite{markovic10,triesch07}

\section*{Acknowledgments}
We thank the file search engine
\href{http://www.findfiles.net}{FindFiles.net}
for support and data collection.
The complete raw data of the Wikipedia/Dmoz
corpus is available for download at
\url{http://www.findfiles.net/public} ~.



\end{document}